\documentclass[prb,twocolumn,superscriptaddress,showpacs,floatfix]{revtex4-2}

\usepackage[utf8]{inputenc}   
\usepackage[british]{babel}   

\usepackage{amsmath}        
\usepackage{amssymb}        
\usepackage{mathtools}      
\usepackage{mathrsfs}       
\usepackage{bm}      
\usepackage{tablefootnote}

\usepackage{braket}
\usepackage{derivative}    

\usepackage{graphicx}      
\graphicspath{{images/}}    
\usepackage{xcolor}        
\usepackage{float}
\usepackage{tikz}
\usepackage{quantikz}

\usepackage{hyperref}

\begin{document}

\title{Chaos and the Approach to the Page Bound in Locally Interacting Spin Chains}
\author{Samuel J. W. Jones}
\email{ed20sjwj@leeds.ac.uk}
\author{M. Basil Altaie}
\author{Benjamin T. H. Varcoe}
\affiliation{School of Physics and Astronomy, University of Leeds, Woodhouse, Leeds LS2 9JT}

\date{\today}

\begin{abstract}
This work considers the role of chaos in the entanglement entropy dynamics of a shrinking subsystem in an Ising-based spin chain evolving unitarily under local dynamics. Two chains, a `system' and an `environment', are coupled to form a pure bipartite state, and the system’s size is dynamically reduced while the environment grows, following the kinematic setup of Page's argument for evaporation. Unitary real-time evolution is simulated by means of matrix product state (MPS) tensor networks, where DMRG and TEBD algorithms are implemented. The entropy rises and falls in the manner of a Page curve. Setting the coupling across the moving boundary to zero reduces the protocol exactly to Page's setting, a fixed set of $N_I$ spins partitioned by a moving cut, so the turnover is then enforced by dimension alone and the amplitude isolates the internal dynamics. At the turnover the chaotic chain reaches approximately $93\%$ of Page's average entropy and the integrable chain approximately $56\%$. How closely the measured entropy approaches the kinematic bound with and without chaos present in the system is probed. To do so integrable (transverse-field) and chaotic (transverse and longitudinal-field) Hamiltonians are compared, classified by adjacent-gap-ratio level statistics, and vary the evaporation interval. It is found that breaking integrability brings the entropy substantially closer to the bound, whereas in the integrable chain it remains well below the bound throughout the evolution.
\end{abstract}
\maketitle

\section{Introduction}
\label{sec.Intro}
\begin{figure}
    \centering
    \includegraphics[width=1\linewidth]{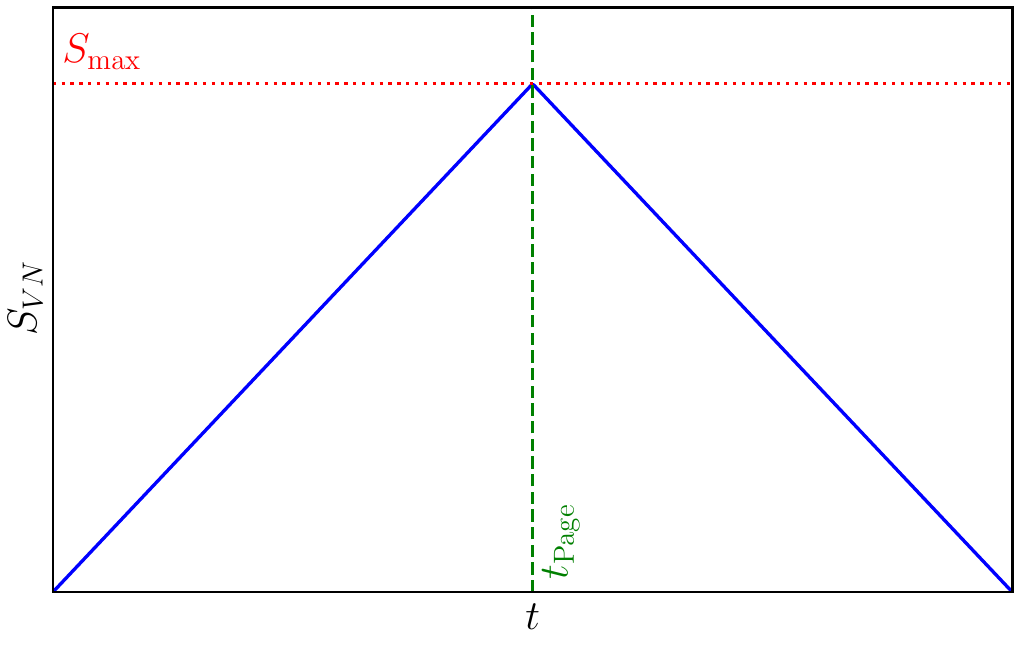}
    \caption{A schematic of the idealised Page curve, depicting the entropy of a shrinking subsystem's complement over time. Entropy grows while the shrinking subsystem carries away degrees of freedom, and falls once its remaining Hilbert space can no longer support further entanglement growth. This turning point is the Page time, $t_{\mathrm{Page}}$.}
    \label{fig:PageCurve}
\end{figure}

Page treated black hole evaporation as a unitary quantum-mechanical process in which a black hole and its surrounding radiation together form a pure bipartite state \cite{Page,page2}. Under this assumption, the entanglement entropy between the two, measured via the von Neumann entropy, rises as Hawking radiation carries degrees of freedom into the environment, then falls once the shrinking black hole can no longer support further entanglement growth \cite{Almheiri2021}. This characteristic rise-and-fall, as depicted in Fig.~\ref{fig:PageCurve}, was derived from a purely kinematic argument about subsystem dimension, largely independent of any microscopic dynamics which the chain may be subject to \cite{Page,page2}.

This kinematic origin motivates a question outside the context of gravity. Page's argument assumes that the state of the shrinking subsystem is typical, so that its entropy approaches the dimensional bound $\min(\log d_A,\log d_B)$, up to a correction of at most half a nat \cite{Page1993avg,PopescuShortWinter2006}. Whether physical dynamics produce such typical states is a separate matter. Chaotic Hamiltonians are expected to do so through eigenstate thermalisation \cite{Deutsch1991,Srednicki1994,Rigol2008,DAlessio2016,VidmarRigol2017}, while integrable ones, whose conserved quantities confine the state to a small part of the Hilbert space, relax instead to a generalised Gibbs ensemble with sub-maximal entanglement \cite{Rigol2007,VidmarHackl2017,Vidmar2018Ising,Bianchi2022}. Using integrable and chaotic variants of the Ising Hamiltonian provides a platform on which to ask how much of its shape is fixed by the schedule of subsystem sizes alone, and how much depends on the internal dynamics.

Existing studies of Page-curve-like behaviour in Hamiltonian systems \cite{Jha2025,Stefan,ray,li,Ares_2022} typically initialise a fermionic chain in a filled state that empties into an environment, so the change in available degrees of freedom is generated by the dynamics itself. Random-circuit models of evaporation \cite{HaydenPreskill2007,Piroli2020} impose the change in subsystem size explicitly, but have no Hamiltonian and so cannot separate integrable from chaotic dynamics. This makes it hard to separate the dimensional constraint from the dynamics whilst also including a longitudinal field which invokes chaos. Here, instead, an explicit, discrete change in the number of degrees of
freedom of each subsystem is imposed, mirroring the dimensional argument in Page's derivation. The process is then exact and identical for every Hamiltonian, any differences between the two chains can be attributed to their internal dynamics and, consequently, to the presence or absence of chaos.

This argument is used to compare an integrable transverse-field Ising model (TFIM) with a chaotic transverse-longitudinal-field (TLFIM) version, diagnosed through the adjacent-gap ratio \cite{OganesyanHuse2007,Atas2013,Gap1}, as well as the effect of a coupling across the moving boundary and of the rate at which the boundary moves. At zero boundary coupling the protocol reduces exactly to Page's setting, so the rise and fall is guaranteed by dimension and serves as a control in which the amplitude depends only on the internal dynamics. The amplitude and late-time behaviour depend strongly on chaos: the entropy approaches the kinematic bound far more closely in the chaotic chain than in the integrable model.

\section{Model}
\label{sec.model}
First, consider the TFIM \cite{Ising1,Ising3,Ising4,Ising5,Ising6}, which is built from solely local interactions and whose dynamics has been realised experimentally in trapped ions \cite{Friedenauer2008,blatt,Zhang2017}, Rydberg arrays \cite{Bernien} and superconducting processors \cite{KimIBM2023}. Under a Jordan-Wigner transformation the TFIM maps to non-interacting fermions, so it is integrable. The TLFIM permits an additional longitudinal field on the TFIM. This has no free-fermion mapping, breaks integrability, and is chaotic over a wide range of parameters \cite{KimHuse2013,AtasBogomolny2017,BCH2011}; Sec.~\ref{sec.gapratio} verifies this for the parameters used here through level statistics. Comparing the two Hamiltonians allows for the separation of the entropy profile that follows from shrinking the subsystem from the part that depends on the internal dynamics.
 
A common choice of Hamiltonian for complex information dynamics \cite{Yasuhiro} is the Sachdev-Ye-Kitaev (SYK) model \cite{SachdevYe1993,SYK}, widely used in holographic settings. It relies on non-local, all-to-all interactions which produce near-maximal information propagation. These dense all-to-all couplings make the model difficult to realise experimentally and to simulate computationally. Such interactions are unnecessary for Page-curve-like entanglement profiles \cite{Jha2025,Stefan,ray,li}, nor are they required for a Hamiltonian to be chaotic. The TLFIM is chaotic with nearest-neighbour interactions only, and chaos does not dictate fast scrambling. This work focuses on the role of chaos and not scrambling, positioning the locally interacting TFIM/TLFIM Hamiltonians as a sufficient choice of spin-model.
 
\begin{figure}
    \centering
    \includegraphics[width=1\linewidth]{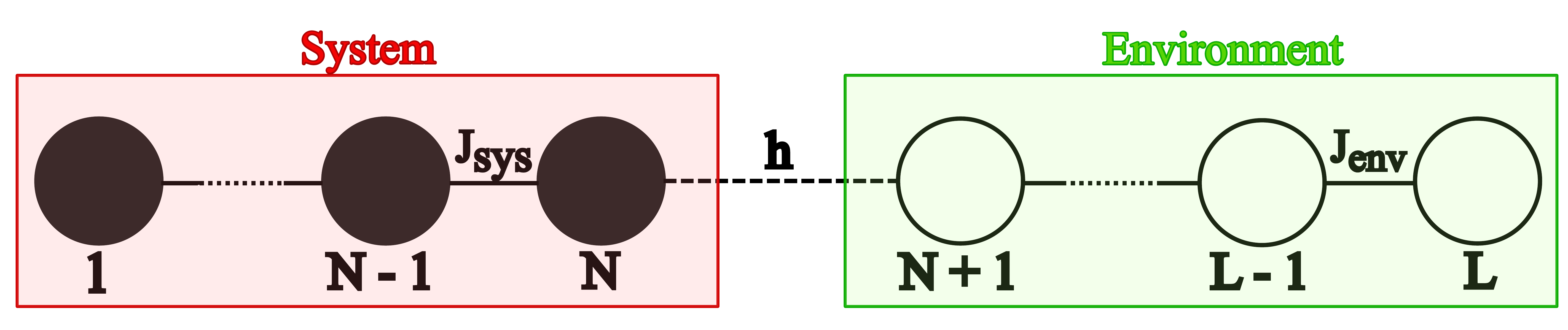}
    \caption{A schematic of the full spin chain modelled by the Hamiltonian given in Eq.~\ref{eq.hamiltonian}. The chain is split, one for the system coloured in red, of size $N$ with interaction strength $J_{sys}$ and one for the environment, in green, with interaction strength $J_{env}$ and has length $M$. The overall size of the system is $L = N+M$ with a Hilbert space of size $\mathcal{H} \sim d^{L}$. The two subsystems are coupled with boundary coupling $h$. To simulate evaporation, the boundary coupling $h$ is  systematically moved to the left to grow the environment and decrease the system whilst allowing the system to evolve through time using a TEBD protocol (see Appendix~\ref{sec.method} for more information) to simulate unitary evolution.}
    \label{fig:chainschematic}
\end{figure}

The chain in this work is bipartite, partitioned into a system of length $N$ and an environment of length $M$, with the sites ordered linearly (Fig.~\ref{fig:chainschematic}). An out-of-equilibrium quench is performed, initialising the system in a non-eigenstate to drive the internal dynamics. To mitigate finite-size boundary effects and to let entanglement leave the system, $M\gg N$, using $N=15$ and $M=100$ as the primary simulation conditions. The chain is governed by the spin Hamiltonian
\begin{equation}
    \label{eq.hamiltonian}
    \hat{H} = \hat{H}_{sys} + \hat{H}_{env} + \hat{H}_{int},
\end{equation}
with terms
\begin{align*}
    \hat{H}_{sys} &= -J_{sys}\sum_{i=1}^{N-1}\hat{S}^z_i\hat{S}^z_{i+1}  -g_{sys}\sum^N_{j=1}\hat{S}^x_j - h_z\sum^N_{j=1}\hat{S}^z_j,\\
   \hat{H}_{env} &= -J_{env}\sum_{i=N+1}^{L-1}\hat{S}^z_i\hat{S}^z_{i+1}  -g_{env}\sum^{L}_{j=N+1}\hat{S}^x_j ,\\
    \hat{H}_{int} &= -h\hat{S}^z_N\hat{S}^z_{N+1},
\end{align*}
$L=N+M$ and $\hat S^\alpha=\hat\sigma^\alpha/2$ are spin-$\tfrac12$ operators (equivalently, in Pauli matrices, the couplings are $J/4$, $g/2$ and $h_z/2$). The case $h_z=0$ yields the standard TFIM and $h_z\neq0$ the TLFIM; the longitudinal field acts on the system only and is not to be confused with the boundary coupling $h$.
 
At $h_z=0$ the chain is critical at $J=2g$. Defining $\lambda=J/2g$, the chain is in its disordered, paramagnetic, phase for $\lambda<1$ and its ordered phase for $\lambda>1$, with ground-state correlation length $\xi=1/|\ln\lambda|$, away from the critical point and a single-particle gap $\Delta=|g-J/2|$. A longitudinal field removes the critical point, so $\lambda=1$ is critical only for the TFIM. Unless stated otherwise it is taken that $J_{sys}=g_{sys}=3$ and $J_{env}=g_{env}=1$, which places both chains in the disordered phase ($\lambda=\tfrac12$, $\xi\approx1.4$ lattice spacings), with gaps $\Delta=1.5$ and $0.5$ respectively. The environment ground state is therefore short-range correlated and gapped. $J_{sys}>J_{env}$ is prescribed such that the two chains have different energy scales, and a boundary coupling $h=\mathcal{O}(J_{sys})$. A pure state $\hat{\rho}$ must satisfy the idempotency and normalisation identities
\begin{equation}
\label{eq.pure identities}
\hat{\rho}^2 = \hat{\rho} \quad \text{and} \quad \text{tr}(\hat{\rho}^2) = 1,
\end{equation}
where $\hat{\rho} = \ket{\psi}\bra{\psi}$. Given that the derivation outlined in \cite{Page, page2} relies on the purity of the total system, both the system and environment subsystems are initialised in pure states. The global initial state is then
\begin{equation}
\label{eq.initial state}\ket{\psi_I} = \ket{\zeta} \otimes \ket{\phi_0},
\end{equation}
where the system is initialised to be
\begin{equation}
\label{eq.zeta}
\ket{\zeta} = \ket{0}^{\otimes \lfloor N/2\rfloor} \otimes \frac{1}{\sqrt{2}}\left( \ket{0101\cdots} - \ket{1010\cdots} \right),
\end{equation}
where the second factor is an antiferromagnetic cat state on the remaining $\lceil N/2\rceil$ spins. For odd $N$ (e.g.\ $N=15$) the product part has $\lfloor N/2\rfloor$ spins and the cat part the remaining $\lceil N/2\rceil$. The environment state $\ket{\phi_0}$ is the state corresponding to $E_0$, satisfying $\hat{H}_{env}\ket{\phi_n} = E_n\ket{\phi_n}$, where $E_0 = \min(\{E_n\})$. This choice of $\ket{\phi_0}$ mimics a vacuum-like environment by minimising internal dynamics.

Two properties of $\ket{\zeta}$ matter for what follows. First, $\ket{\psi_I}$ is a product state across the system–environment cut, so $S_{VN}(0)=0$, matching the pure initial state of Page's argument; all entanglement between system and environment is generated by the dynamics. The cat state carries at most $\ln 2$ of entanglement across any internal cut, as does a GHZ state \cite{ghz}, so the system itself also starts close to a product state.

Second, $\ket{\zeta}$ lies close to the centre of the many-body spectrum, which is where Eq.~\ref{eq.bound S} is the appropriate reference. The transverse field has zero expectation value in $\ket{\zeta}$, the cat part has zero longitudinal magnetisation, and its antiferromagnetic bonds cancel most of the ferromagnetic bonds of the product part, giving
\begin{equation}
\langle\hat H_{sys}\rangle = \frac{J_{sys}}{4}\left(\lceil N/2\rceil-\lfloor N/2\rfloor\right) - \frac{h_z}{2}\lfloor N/2\rfloor ,
\label{eq.initial energy}
\end{equation}
that is $J_{sys}/4-7h_z/2$ for $N=15$, against the infinite-temperature value $\mathrm{tr}\,\hat H_{sys}/2^N=0$. Exact diagonalisation of $\hat H_{sys}$ at $N=15$ gives the energy density $\epsilon=(\langle\hat H_{sys}\rangle-E_{\min})/(E_{\max}-E_{\min})$ listed below, together with the offset from the spectral centre in units of the infinite-temperature energy width, $z=\langle\hat H_{sys}\rangle/\sigma_\infty$ with $\sigma_\infty^2=\mathrm{tr}\,\hat H_{sys}^2/2^N$:
\begin{center}
\begin{tabular}{lcccc}
\hline\hline
$h_z$ & 0 & 1.5 & 3 & 6 \\
$\epsilon$ & 0.516 & 0.458 & 0.429 & 0.382 \\
$z$ & $+0.12$ & $-0.64$ & $-1.12$ & $-1.52$ \\
\hline\hline
\end{tabular}
\end{center}
A thermalising system at the centre of its spectrum approaches infinite temperature, and hence the typical-state value $N(t)\ln 2$ \cite{DAlessio2016,VidmarRigol2017}; at lower energy density it would instead thermalise to a lower entropy, and the appropriate comparison would be the thermal entropy at that energy. The chaotic chain ($h_z=1.5$) starts further from the spectral centre than the integrable one, so the difference between them reported below is not produced by a more favourable initial energy.

Standard unitary evolution with static subsystem sizes fails to capture the loss of degrees of freedom characteristic of Hawking radiation, the key driver in the production of the Page curve. In physical evaporation, the continuous release of radiation reduces the black hole's mass and the dimension of its associated Hilbert space. To model this, albeit in a significantly smaller regime, a dynamic boundary condition that discretely transfers spins from the system to the environment is implemented. The time-dependent subsystem lengths, $N(t)$ and $M(t)$, are governed through
\begin{align}
\label{eq.NM floor} 
N(t) = N_I - \left\lfloor \frac{t}{T} \right\rfloor \text{ and }
M(t) = M_I + \left\lfloor \frac{t}{T} \right\rfloor,
\end{align}
where $T$ represents the evaporation interval. This setup ensures that the system shrinks only at discrete steps $t = nT$ for $n \in \mathbb{Z}$. The resulting time-dependent Hamiltonian is then
\begin{align*}
    \hat{H}(t) = &-J_{sys}\sum_{i=1}^{N(t)-1}\hat{S}^z_i\hat{S}^z_{i+1}  -g_{sys}\sum^{N(t)}_{j=1}\hat{S}^x_j - h_z\sum^{N(t)}_{j=1}\hat{S}^z_j\\
  &-J_{env}\sum_{i=N(t)+1}^{L-1}\hat{S}^z_i\hat{S}^z_{i+1}  -g_{env}\sum^{L}_{j=N(t)+1}\hat{S}^x_j\\ &-h\hat{S}^z_{N(t)}\hat{S}^z_{{N(t)}+1}.
\end{align*}
While $\hat{H}(t)$ is globally time-dependent, the Hamiltonian remains piecewise constant between evaporation events. This allows for standard unitary propagation within each interval $T$, followed by a discrete update of the subsystem boundaries. The simulation of large-scale quantum systems is constrained by the exponential scaling of the Hilbert space, $\mathcal{H} \sim d^{L}$. Direct diagonalisation of $\hat{H}_{env}$ becomes intractable as $M$ increases. To circumvent this dimensionality problem, MPS is utilised, a tensor network formalism that decomposes the high-order state tensor into a chain of low-rank local tensors \cite{Orus, vidal}.

\section{The Page Curve}
\label{sec.page}
The rise and fall of the entropy profile is expected from geometric arguments. The entropy starts near zero because the initial state is a product across the system-environment cut, and it is capped by the dimension of the smaller system, which falls to zero as the last spin leaves. With $N=15$ and $M=100$, the two subsystems are never comparable in size, hence the turnover is not set by $d_A=d_B$. At $h=0$ the system can only entangle with spins that have already left it, so the turnover is expected near $N(t)\approx N_I/2$, where the remaining system and the evaporated spins have equal dimension. For $h\neq0$ the system can also entangle with the original environment, and the peak instead occurs where the growing entropy meets the descending bound $N(t)\ln 2$. In both cases the observed entanglement behaviour is consistent with the dimensional constraints of a unitary bipartite model undergoing subsystem reduction. During the early stages, the system is dominated by information transfer, characterised by the interaction strengths and the dynamic coupling $h$, which determines the rate of entanglement growth. Conversely, after the Page time, the system size dominates the environment entropy. The limited degrees of freedom in the system restrict the entropy in accordance with von Neumann behaviour for a pure bipartite model where
\begin{equation}
\label{eq.Svn}
S_{env} = -\text{tr}_{env}(\hat{\rho}_{env} \ln{\hat{\rho}_{env}}) \equiv S_{sys}.
\end{equation}
The rates of growth and decay need not be equal, and it is shown that this is dependent on the internal dynamics. At $t=0$, the environment begins in its ground state with its associated entanglement. At the final time-step when $N=0$, the simulation does not record the entropies. This is an artifact of measuring the von Neumann entropy just before an evaporation event. After the final spin leaves the system, the entropy is zero by definition, as seen from Eq.~\ref{eq.Svn} since $\hat{\rho}_{sys} = 1$. Thus, the entropy begins and ends at zero, conserving information. The theoretical maximum of the system, bound by the smaller dimension of the two subsystems is given as
\begin{equation}
\label{eq.bound S}
    S_A = \text{min}(\text{ln}(d_A),\text{ln}(d_B)).
\end{equation}
This value is approached by typical states \cite{Page1993avg,Bianchi2022}, so it is expected to be approached once the entropy has peaked. It is not expected that the entropy follow this behaviour in the first half of the evolution since the initial state is pure. The system is always bound by its own dimensionality rather than the environment size since it always retains the smaller dimensions. Fig.~\ref{fig:hzplot} confirms the integrable chain does not follow the bound. In this case, the entropy rises and falls, but remains well below the bound throughout the second half of the evolution. This is not surprising for the TFIM. The integrable model maps to free fermions, and the conserved quantities heavily restrict the areas of the Hilbert space which the dynamics can reach, so there is no reason for the evolving state to be typical. The Page-type bound could instead be approached by dynamics that mix generic states, which may be done by breaking integrability, challenging whether the entropy then follows the bound more faithfully.

\subsection{The Adjacent Gap Ratio}
\label{sec.gapratio}
 For the sorted eigenvalues $E_n$ within a symmetry sector, with gaps $\delta_n=E_{n+1}-E_n$, the adjacent gap ratio is
\begin{equation}
    r_n=\frac{\min(\delta_n,\delta_{n+1})}{\max(\delta_n,\delta_{n+1})}\in[0,1].
\end{equation}
Taking a ratio of neighbouring gaps cancels the local density of states, so no unfolding of the spectrum is needed \cite{OganesyanHuse2007,Atas2013}. Levels of integrable systems are uncorrelated and cluster, giving small $r_n$ and, for Poisson statistics, $\langle r\rangle=2\ln2-1\approx0.386$. Levels of chaotic systems repel one another, giving similar neighbouring gaps, for the Gaussian orthogonal ensemble, appropriate to this real symmetric Hamiltonian, $\langle r\rangle\approx0.531$ \cite{BGSW1984,Atas2013}.

The ratio is only meaningful within a symmetry sector, since an unresolved symmetry superposes independent spectra and biases $\langle r\rangle$ towards the Poisson value \cite{Gap1}. The open chain with uniform couplings is invariant under spatial reflection for all $h_z$, and under spin-flip parity when $h_z=0$. These symmetries are resolved before calculating the statistics. The system Hamiltonian is diagonalised, Eq.~\ref{eq.hamiltonian}, on its own (no boundary motion) for $N=12$ spins and average $r_n$ over the central third of each sector, with the standard error of the mean. For $J=g=3$ it is found
\begin{center}
\begin{tabular}{lcccc}
\hline\hline
$h_z$ & 0 & 1.5 & 3 & 6 \\
$\langle r\rangle$ & 0.312 & 0.525 & 0.485 & 0.406 \\
$\sigma_{\langle r\rangle}$ & 0.007 & 0.007 & 0.007 & 0.008 \\
\hline\hline
\end{tabular}
\end{center}
At $h_z=1.5$ the ratio is consistent with GOE, so it is taken as the chaotic case. At $h_z=0$ the TFIM is far from it. Its many-body levels are sums of single-particle energies and need not follow Poisson statistics, and $\langle r\rangle$ lies below the Poisson value; what matters here is the
separation from the chaotic value. Larger fields give an intermediate ratio \cite{Gap3}, so the longitudinal field does not increase chaoticity monotonically.

\subsubsection{Magnetisation Tracking}
\begin{figure*}
        \centering
    \includegraphics[width=1\linewidth]{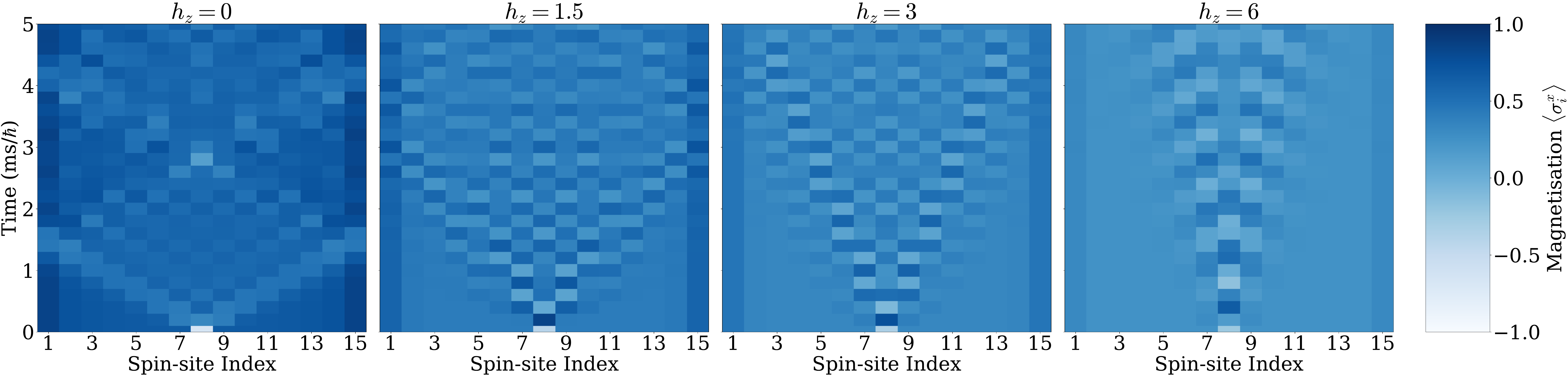}
    \caption{Local-quench dynamics of the spin chain for $N = 15$, $J = g = 3$ and the values of $h_z$ indicated above each panel. The system starts in the ground state of $H$ at that $h_z$, with $\sigma^z$ applied to spin site 8, and is evolved exactly under the same $H$ up to $t=5$. Each panel shows $\langle\sigma^x_i\rangle$ as a function of spin site index (horizontal axis) and time $t$, with $\hbar=1$ (vertical axis, increasing upwards). The colour scale is common to all panels.}
    \label{fig. Magnetisation tracking}
\end{figure*}
To qualitatively illustrate how these spectral differences manifest in the dynamics, an observable is tracked through time. Fig.~\ref{fig. Magnetisation tracking} shows the evolution of $\langle\sigma^x_i\rangle$ after a local quench for $h_z \in [0,1.5,3,6]$ \cite{Jurcevicetal2014}. At $h_z = 0$ the perturbation spreads outward in a well-defined front and the magnetisation near the ends of the chain remains large throughout. At $h_z = 1.5 \text{ and } 3$ the front is followed by an irregular bulk that settles around a roughly uniform value. At $h_z = 6$ the perturbation stays confined closer to the quench site and the outer spins change comparatively little when subject to the same evolutionary time frame. For this larger field it is found $\langle r \rangle = 0.406 \pm 0.008$, intermediate between the Poisson and GOE values, so the suppressed spreading coincides with a departure from the GOE statistics. These observations are qualitatively consistent with the level-statistics results, including the non-monotonic dependence on $h_z$. However, a single quench cannot distinguish chaotic relaxation from de-phasing in a nearly integrable system, though it is an indicator of chaos that $h_z=1.5$ appears to explore more states in the given time frame compared to the other established values. Since entanglement in Ising chains grows linearly in time after a quench in both the integrable and the non-integrable case \cite{calabrese2005,KimHuse2013}, the growth law itself does not distinguish the two chains. What breaking integrability is expected to change is the value at which the entropy saturates, thermal for a chaotic chain \cite{DAlessio2016,KimHuse2013} and set by the generalised Gibbs ensemble for an integrable one \cite{AlbaCalabrese2017}, and it is this that is compared with the bound.

\subsection{The role of Chaos in $h_z$ sweeps}
\label{sec. hz}
\begin{figure}
    \centering
    \includegraphics[width=1\linewidth]{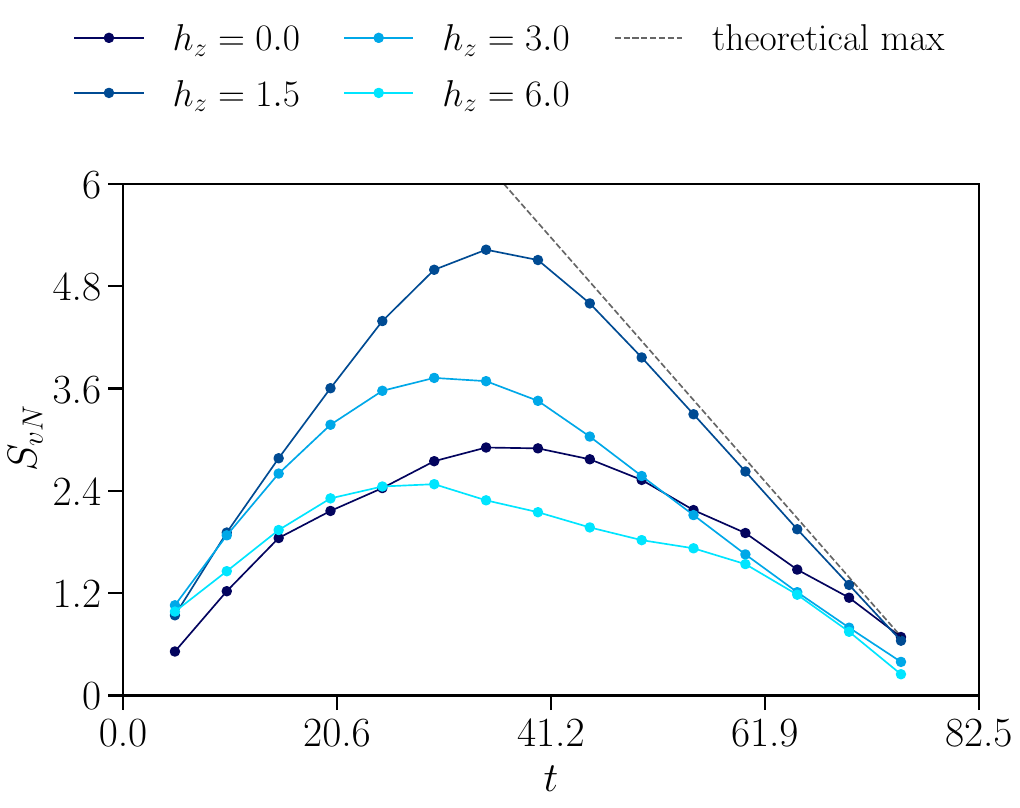}
    \caption{von Neumann entropy of the environment under the evaporation protocol for $h_z = 0, 1.5, 3 \text{ and }6$, with $N = 15$, $M = 100$, $T = 5$ and $h = 3$. Markers are the measurements made just before each evaporation event. The dashed line is the bound of Eq. \ref{eq.bound S}, where $S_{VN}=N(t)\ln 2$. Breaking integrability brings the entropy closer to the bound.}
    \label{fig:hzplot}
\end{figure}
It has been established that the level statistics identify $h_z = 1.5$ as the chaotic case. The evaporation protocol is run for the four $h_z$ fields with varying gap ratios in Sec.~\ref{sec.gapratio}; the results are shown in Fig.~\ref{fig:hzplot}. Since $M \gg N$, the bound of Eq.~\ref{eq.bound S} reduces to $N(t)\ln 2$, shown as the dashed line. All curves rise and fall, as the geometric argument requires, but they differ in how closely they follow the bound. At $h_z = 0$ the entropy peaks at $S_{VN} \approx 2.9$. At $h_z = 1.5$ the entropy grows faster, peaks at $S_{VN} \approx 5.2$ near $t \approx 35$, and for $t \gtrsim 45$ its descending branch lies close to the bound, so it is close to the idealised profile of Fig.~\ref{fig:PageCurve}. The $h_z=3$ curve lies between the two, peaking at $S_{VN}\approx3.7$, while the $h_z=6$ curve lies below the integrable one, peaking at $S_{VN}\approx2.5$. The ordering of the $h_z=0$, $1.5$ and $3$ curves follows $\langle r\rangle$, but $h_z=6$ does not, its gap ratio is larger than at $h_z=0$, yet its entropy is lower. At this field the longitudinal term dominates the Hamiltonian ($h_z/2=3$ against $g/2=1.5$ and $J/4=0.75$ in Pauli units), so the dynamics approaches the trivially polarised large-field limit, consistent with the confinement of the local perturbation in Fig.~\ref{fig. Magnetisation tracking}, as expected when a longitudinal field confines domain-wall excitations \cite{kormos2017}, and the drift of $\langle r\rangle$ back towards the Poisson value. The initial state is also colder at this field ($\epsilon=0.38$, $z=-1.52$, Sec.~\ref{sec.model}), but this accounts for little of the deficit: for an approximately Gaussian density of states the thermal entropy of the full system lies $z^2/2\approx1.2$ nats below $N\ln 2$, and that of a subsystem of half the chain correspondingly less, far short of the $\approx2.5$-nat gap between the $h_z=6$ and $h_z=1.5$ peaks. The ordering of the peaks at $h_z=0$ and $1.5$ is consistent with the level statistics. The intermediate fields do not follow $\langle r\rangle$ monotonically: $h_z=6$ has a larger $\langle r\rangle$ than $h_z=0$ but a lower peak. With four values of $h_z$ and a single initial state the conclusion may be drawn only that breaking integrability, in the sense of a clear departure from the integrable level statistics, brings the entropy closer to the bound, and not a quantitative relation between $\langle r\rangle$ and the entropy profile.

\subsection{The role of the coupling $h$}
\label{sec.h}
\begin{figure}
    \centering
    \includegraphics[width=1\linewidth]{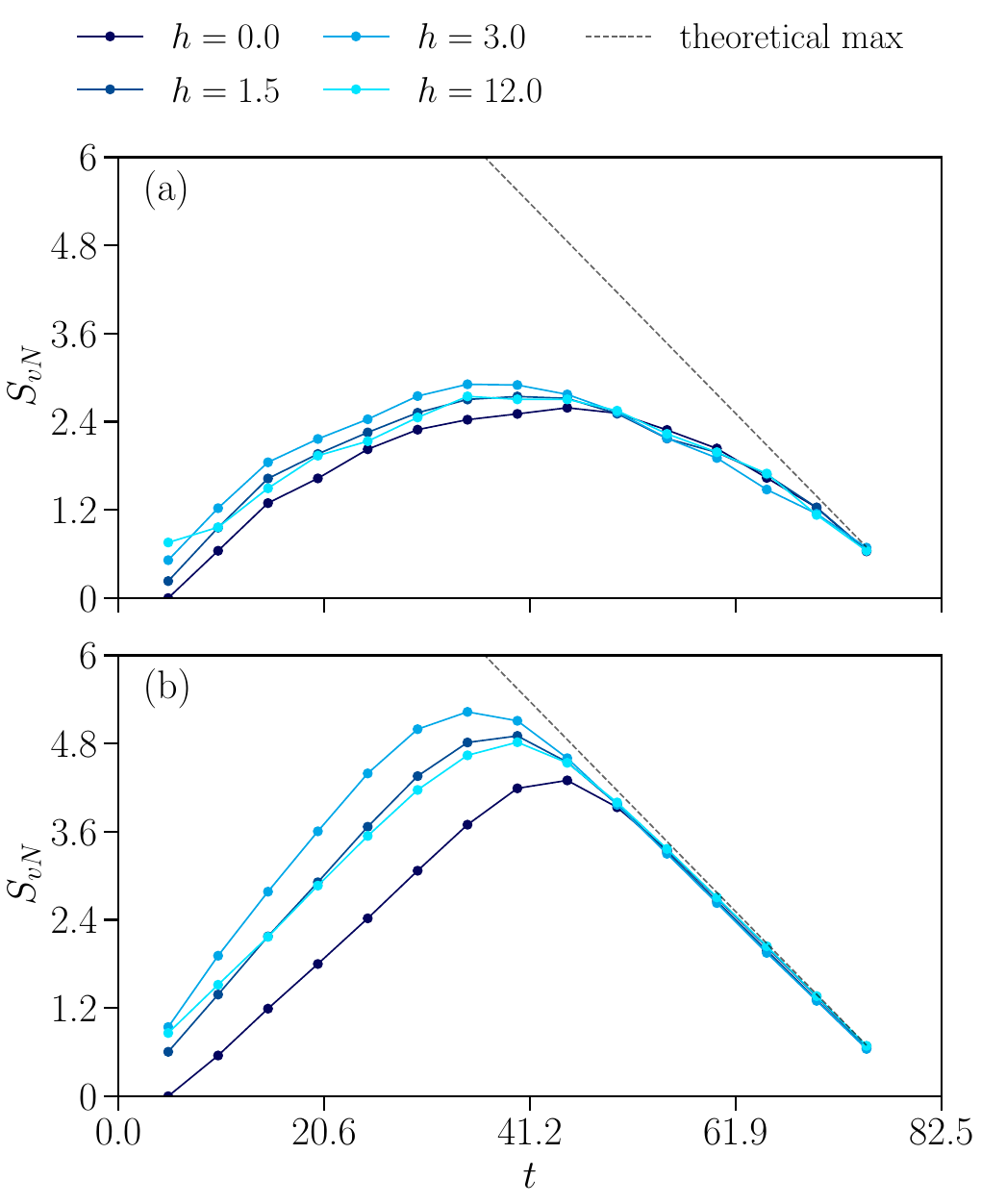}
    \caption{von Neumann entropy of the environment under the evaporation protocol for a sweep through $h$. $(a)$ is the integrable system with $h_z=0$ and $(b)$ is the chaotic system with $h_z =1.5$. The other parameters are $N = 15$, $M = 100$ and $T = 5$. Markers are the measurements made just before each evaporation event. The dashed line is the bound of Eq.~\ref{eq.bound S}, where $S_{VN}=N(t)\ln 2$. For $h=0$ the relevant references are the ceiling $\min(N(t),N_I-N(t))\ln 2$ and Page's average entropy $\ln d_A-d_A/2d_B$ for $N_I$ spins \cite{Page1993avg}.}
    \label{fig:hplot}
\end{figure}
At $h=0$ the remaining system never interacts with the environment, and unitaries acting only on the complement leave $S_{VN}$ unchanged. The protocol is then exactly Page's: a pure state of $N_I$ spins, partitioned into $N(t)$ and $N_I-N(t)$ spins. The rise and fall is therefore guaranteed by the ceiling $\min(N(t),N_I-N(t))\ln 2$, and this case is used as a control in which the amplitude depends only on the system's internal dynamics. Page's average for a random state, $S\approx\ln d_A-d_A/2d_B$ \cite{Page1993avg}, gives $\approx 4.60$ at the turnover; the chaotic chain reaches $S_{VN}\approx4.3$ and the integrable chain $S_{VN}\approx2.6$ (Fig.~\ref{fig:hplot}). Any dependence on $h$ beyond this ceiling comes from the environment taking part in the dynamics. In the integrable chain, increasing $h$ causes only a slight change in the output curve and the entropy remains well below the bound after the peak. In the chaotic chain, increasing $h$ from $0$ raises the peak to $S_{VN}\approx5.2$ at $N=8$, and the descending branch then follows the bound closely. This is interpreted as the coupling allowing the environment to participate in the entanglement dynamics, which the closed ceiling above cannot account for. Increasing $h$ from $3$ to $12$ gives a similar curve, so the peak is insensitive to $h$ above about $3$. This dependence on $h$ is seen only in the chaotic chain, for a single initial state and parameter set.

\section{The Evaporation Interval $T$}
\label{sec:T}
\begin{figure}
    \centering
    \includegraphics[width=1\linewidth]{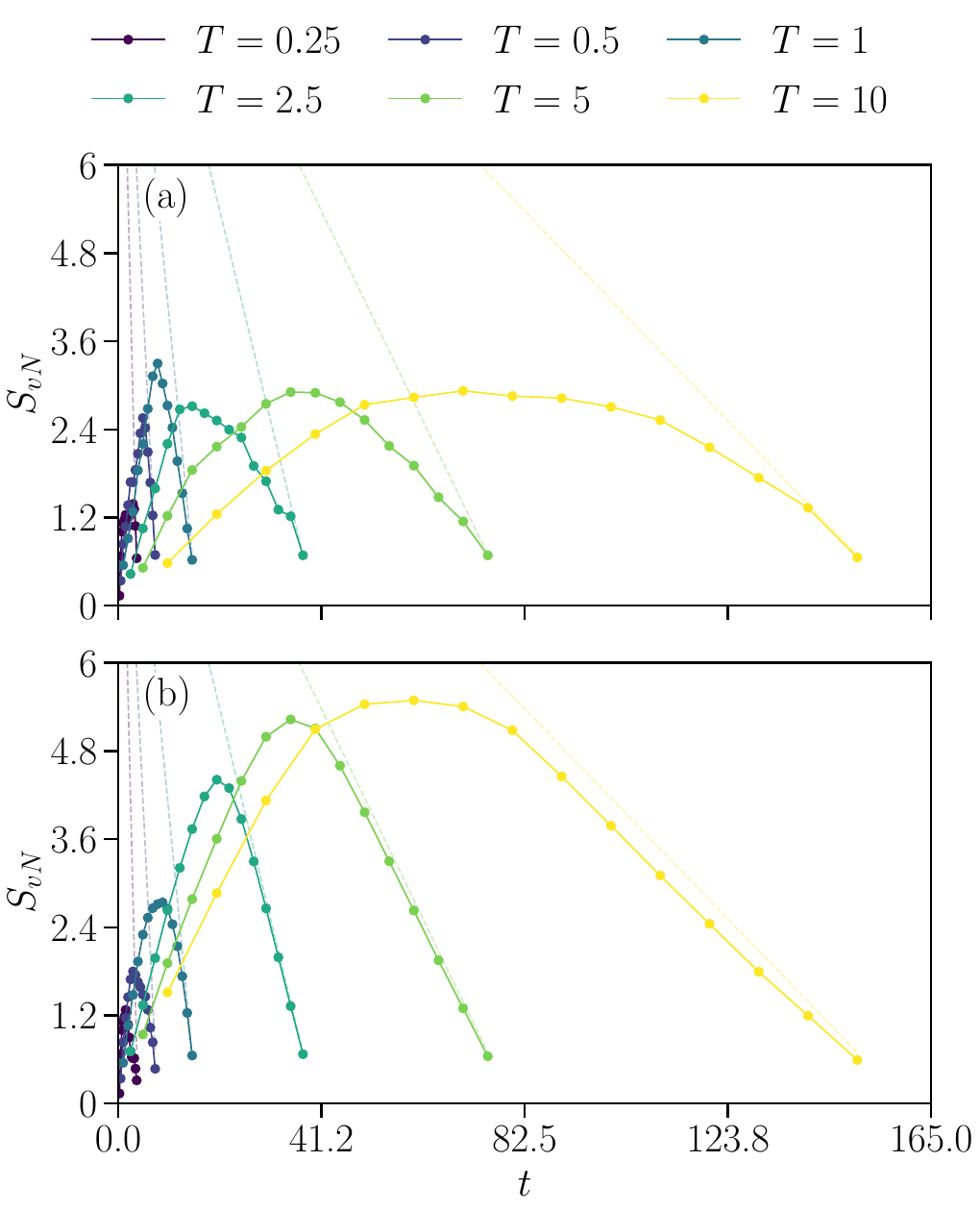}
    \caption{von Neumann entropy of the environment under the
evaporation protocol for various evaporation intervals $T$, with $N = 15$. Markers are the measurements
made just before each evaporation event. The dashed lines
are the bound of Eq.~\ref{eq.bound S}, $S_{VN} = N(t) \ln 2$, colour-coded with each $T$. The competition between $T$ and the information propagation speed is present and clear in small $T$ protocols. }
    \label{fig:Tplot}
\end{figure}
The boundary advances by one spin every interval $T$, it moves at speed $v_b = 1/T$ lattice spacings per unit time. One expects the entropy profile to depend on how $v_b$ compares with the speed at which entanglement spreads through each chain. This is tested by varying $T$ at fixed $N = 15$, $M = 100$ and
$h = 3$, for $h_z = 0$ and $h_z = 1.5$. For the isolated TFIM, written in Pauli form as $\hat H = -J'\sum_i \hat Z_i\hat Z_{i+1} - g'\sum_i \hat X_i$ with $J' = J/4$ and $g' = g/2$, the Jordan--Wigner transformation gives free fermions with dispersion $\varepsilon(k) = 2\sqrt{J'^2 + g'^2 - 2J'g'\cos k}$ \cite{Ising4,CEF2011}, whose maximal group velocity is
\begin{equation}
  v_{\max} = \max_k \varepsilon'(k) = 2\min(J', g').
  \label{eq:vmax}
\end{equation}
See Ref.~\cite{CEF2011} for this dispersion result. As such, entanglement travels through the system at speeds up to $v_{\max}$ \cite{calabrese2006,calabrese2005}. For the system chain ($J_{sys} = g_{sys} = 3$) Eq.~(\ref{eq:vmax}) gives $v_{sys} = 1.5$, and for the environment chain ($J_{env} = g_{env} = 1$) $v_{env} = 0.5$. The boundary speed equals these at $T_{sys} = 1/v_{sys} \approx 0.67$ and $T_{env} = 1/v_{env} = 2$. Of the values used, $T = 0.25$ and $0.5$ move the boundary faster than both chains, $T = 1$
faster than the environment only, and $T \geq 2.5$ slower than both. Eq.~(\ref{eq:vmax}) is exact only for the homogeneous isolated chain at $h_z = 0$. The chain under scrutiny is inhomogeneous, with $J_{sys} \neq J_{env}$ and a boundary coupling, and for $h_z \neq 0$ it is non-integrable, so the quasiparticle picture is not exact; $v_{sys}$ and $v_{env}$ are used as reference scales. The Lieb--Robinson velocity \cite{lieb1972} bounds the true spreading speed but is generally not reached in systems \cite{cheneau2022}, and in non-integrable systems entanglement grows ballistically \cite{KimHuse2013}. Therefore it is not expected to be a perfectly sharp crossover at $T_{sys}$ or $T_{env}$. 

Fig.~\ref{fig:Tplot} shows the competition between the entropy growth and the system resizing events through the entropy profile for $T=0.25$ to $10$ at $h=3$. When the boundary moves faster than entanglement spreads through either chain ($T\lesssim0.5$), the peak entropy is low and qualitatively similar in the integrable and chaotic chains ($S_{VN}\approx1.3$--$1.8$). The internal dynamics have little time to act, which limits the entropy growth, so the bound cannot be reached. As $T$ increases the chaotic peak rises smoothly, reaching $S_{VN}\approx5$ for $T\gtrsim5$, while the integrable peak stays near $2.9$ and is non-monotonic in $T$, with a larger peak at $T=1$ ($\approx3.3$). The rise with $T$ is gradual, with no sharp crossover at $T_{\rm sys}$ or $T_{\rm env}$, consistent with the quasiparticle picture being only approximate here. The difference between the chains therefore appears only when the boundary moves slowly compared with the dynamics, and $T=5$ lies in this regime. For large $T$, i.e. $T=10$ in the chaotic regime the entropy is saturated to the entropy bound constrained by $N(t)$, hence the long-time evolution does not facilitate further growth of entropy in this case.

\section{Conclusions}
\label{sec.disc}
This work has explored the role of chaos within a local Ising spin model by comparing how closely the entanglement entropy approaches the kinematic bound $S_A = \min(\ln d_A, \ln d_B)$ under an integrable and a chaotic regime, undertaking a deterministic subsystem-resizing protocol. MPS techniques are employed to simulate a system–environment chain undergoing subsystem resizing, while tracking the entropy at each resizing event. At zero boundary coupling the protocol reduces exactly to Page's setting, where the rise and fall is guaranteed by dimension; there, the chaotic chain reaches close to Page's average entropy at the turnover while the integrable chain remains far below it. Breaking integrability with a longitudinal field brings the entropy amplitude substantially closer to the bound, whereas in the integrable chain it remains well below the bound throughout the second half of the evolution. Chaos is diagnosed by the adjacent gap ratios of the energy levels, a standard probe that distinguishes the chaotic and integrable regimes of the Hamiltonian through the distribution of the energy levels.

In the literature, SYK Hamiltonians are often employed to simulate black hole information dynamics due to its fast-scrambling properties. In contrast, the model presented here utilises the TLFIM, which is constrained by local interactions and the Lieb-Robinson velocity, whilst the chaos is introduced via a longitudinal field which is not present in the standard TFIM. In these simulations, nearest-neighbour chaos is enough to bring the entropy substantially closer to the bound, so fast scrambling is not required for this effect. The boundary coupling $h$ matters only in the chaotic chain, where a non-zero $h$ raises the peak entropy above its $h=0$ value; in the integrable chain the dependence on $h$ is weak and plays no significant role in the outcomes. 

A possible advantage of the TLFIM is its compatibility with current quantum architectures. While ground-state initialisation can be explored on quantum annealers \cite{Ising8, Pelofske}, the dynamic evaporation process is well-suited for gate-based quantum computers, such as those developed by IBM \cite{KimIBM2023,tindall}. Specifically, the Trotter-Suzuki decomposition employed in the Time-Evolving Block Decimation (TEBD) simulations translates directly into sequences of one- and two-qubit gate operations. The locality of the interactions minimises the requirement for SWAP gates during the implementation of the TEBD protocol, thereby reducing a primary source of gate noise and decoherence that would otherwise plague non-local models. The entanglement entropy could be measured on such hardware (Fig.~\ref{fig: Toymodel}) via randomised measurements of the second R\'enyi entropy \cite{Elben2018,Brydges2019}; implementing this for the evaporation protocol is left to future work.

\section*{Acknowledgements}
We would like to thank Jiannis K. Pachos for his insight into
black hole simulations. S.J. would also like to acknowledge Ryan Smith for his knowledge and support on MPS tensor networks and Max Davies for many stimulating discussions on this work. This work was partially supported by the EPSRC grant UKRI1337: Anyons24.
\appendix
\section{Methodology}
\label{sec.method}
\begin{figure}
     \centering
     \includegraphics[width=1\linewidth]{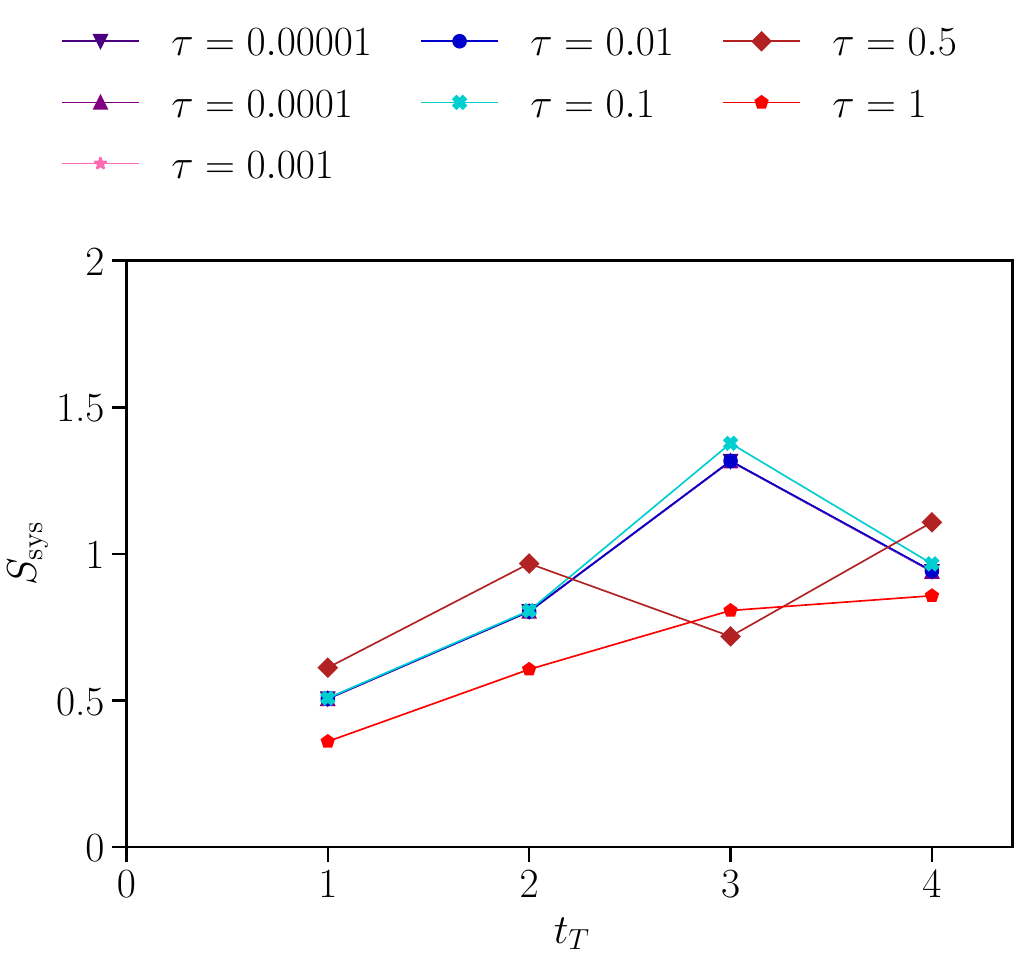}
     \caption{A simulation of the time evolution of the time independent Hamiltonian $\hat{H}$ through unitary evolution simulated with a TEBD protocol for a variable simulation cut-off (see Appendix \ref{sec.method} for more information). The x-axis denotes the time evolution passed, simulated in increments of $\tau$ with an evaporation time-step $T = 5$, each marker denotes a point where an evaporation event was performed. When $\frac{t}{T} \in \mathbb{Z}$  the von Neumann entropy is measured, recorded and then an evaporation event is performed. $N=3$, $M=7$ and $h = 3$ are kept constant. This is done to ensure $\tau = 0.01$ is a viable choice of time step to simulate the evolution.}
     \label{fig:TEBD}
 \end{figure}

The following method was implemented using ITensor.jl \cite{itensor}. The initial state is designated as Eq.~\ref{eq.initial state}. This is done by creating both $\ket{\zeta}$ on sites $1 \text{ to } N$ and $\ket{\phi_0}$ on sites $N+1 \text{ to } L$ separately as MPS, and then introducing a bond of dimension $1$ to connect them. For $\ket{\zeta}$ a product-state MPS is created with the first $\lfloor N/2\rfloor$ sites in $\ket{0}$ and the remaining sites in $\ket{1}$, apply a Hadamard gate to the first of the $\ket{1}$ sites and then a chain of $CNOT$ gates between neighbouring sites, which produces the antiferromagnetic cat state of Eq.~\ref{eq.zeta}. To create the ground state $\ket{\phi_0}$ it must first be known what it actually is. Analytically, the ground state is found by exact diagonalisation of the Hamiltonian matrix, however given the size of the matrix $H_{env}$ scales exponentially with the Hilbert space, this becomes computationally intractable very quickly. Since the Hamiltonian is a sum of local terms, the ground state may be efficiently approximated through the Density Matrix Renormalisation Group Algorithm (DMRG) \cite{white1, white2,scholl,SCHOLL2,DAVIDSON}. This is a variational algorithm which exploits the idea that certain ground states (including those associated with the TLFIM) occupy a very small area of the full Hilbert space. The DMRG calculation was converged over $10$ sweeps, achieving a maximum allowed bond dimension of $\chi = 100$ and a final truncation error of $\epsilon < 10^{-10}$. A convergence of the ground state at low bond dimension is noted circa $\chi = 40$, this ensures the maximum allowed bond dimension of $\chi = 100$ is well suited to an accurate approximation of the ground state. To simulate real-time evolution a second algorithm: a TEBD algorithm \cite{vidal,white3,suzuki} is implemented. Starting from the initial state $\ket{\psi_I}$, a second-order Trotter-Suzuki decomposition \cite{vidal,Vidal2} of the standard time independent evolution operator is used. It is permissible to not use the time dependent Trotter-Suzuki algorithm \cite{Hatano} since between evaporation events the Hamiltonian is time independent. In this case, the Trotter step was chosen to be $\tau = 0.01$ with a cut-off of $\epsilon = \tau^2 \cdot 10^{-6}$. The maximum bond dimension was $\chi_{max} = 300$. At regular intervals when $\frac{t}{T} \in \mathbb{Z}$, where $T\gg \tau$, the von Neumann entropy of the environment system is calculated. The von Neumann entropy is given by
\begin{equation}
    S_{VN} = -\sum_i \lambda_i\ln(\lambda_i),
\end{equation} where $\lambda_i$ are the eigenvalues of the reduced density matrix found via the Schmidt decomposition of the environment. After this calculation, the lengths of each chain are redefined, such that $N \rightarrow N-1$ and $M \rightarrow M+1$, introducing a subsystem resizing into this system. This is computationally realised by adjusting the Hamiltonian such that $H_{sys}$ acts on one less spin site and $H_{env}$ acts on one more spin site for the next evolution interval of $T$. The spins which are changed are the ones at the boundary, thus moving the dynamic coupling to the left in Fig.~\ref{fig:chainschematic}, decreasing the system. The entropy at the end of the next interval is calculated in accordance with the new size of the environment. The total chain itself remains constant in its size of $L$.
 \label{sec.TEBD}
 In Fig.~\ref{fig:TEBD} it can be seen that $\tau = 0.01$ provides a good approximation to the time evolution without becoming computationally too expensive to run for the fixed cut-off. A larger value, $\tau =0.1$, gives qualitatively the same results at lower computational cost for this cut-off. The truncation cut-off is taken proportional to $\tau^2$, so that it scales consistently with the Trotter error across all sweeps. Convergence is seen as $\tau \rightarrow 0$, with no visible change below $\tau = 0.01$.

    \begin{figure*}
        \centering
        \resizebox{\linewidth}{!}{%
            \begin{quantikz}
                \lstick{$\ket{0}$}&&&\slice{}&\gate[1]{R_x(\phi_{sys})}&\gate[1]{R_z(\theta_z)}&\gate[2]{R_{zz}(\theta_{sys})}&&\gate[1]{R_x(2\phi_{sys})}&\gate[1]{R_z(\theta_z)}&\gate[2]{R_{zz}(\theta_{sys})}&&\gate[1]{R_x(\phi_{sys})}&\ldots \\
                \lstick{$\ket{0}$}&&&&\gate[1]{R_x(\phi_{sys})}&\gate[1]{R_z(\theta_z)}&&\gate[2]{R_{zz}(\theta_{sys})}&\gate[1]{R_x(2\phi_{sys})}&\gate[1]{R_z(\theta_z)}&&\gate[2]{R_{zz}(\theta_{sys})}&\gate[1]{R_x(\phi_{sys})}&\ldots \\
                \lstick{$\ket{1}$}&\gate[1]{H}&\ctrl{1}&&\gate[1]{R_x(\phi_{sys})}&\gate[1]{R_z(\theta_z)}&\gate[2]{R_{zz}(\theta_{sys})}&&\gate[1]{R_x(2\phi_{sys})}&\gate[1]{R_z(\theta_z)}&\gate[2]{R_{zz}(\theta_{sys})}&&\gate[1]{R_x(\phi_{sys})}&\ldots\\
                \lstick{$\ket{1}$}&&\targ{}&&\gate[1]{R_x(\phi_{sys})}&\gate[1]{R_z(\theta_z)}&&\gate[2]{R_{zz}(\theta_{h})}&\gate[1]{R_x(2\phi_{sys})}&\gate[1]{R_z(\theta_z)}&&\gate[2]{R_{zz}(\theta_{h})}&\gate[1]{R_x(\phi_{sys})}&\ldots \\
                \lstick{$\ket{0}$}&\gate[1]{H}&\ctrl{1}&\ctrl{1}&\gate[1]{R_x(\phi_{env})}&&\gate[2]{R_{zz}(\theta_{env})}&&\gate[1]{R_x(2\phi_{env})}&&\gate[2]{R_{zz}(\theta_{env})}&&\gate[1]{R_x(\phi_{env})}&\ldots \\
                \lstick{$\ket{0}$}&&\gate[1]{R_y(\eta_I)}&\targ{}&\gate[1]{R_x(\phi_{env})}&&&&\gate[1]{R_x(2\phi_{env})}&&&&\gate[1]{R_x(\phi_{env})}&\ldots
            \end{quantikz}%
        }
        \caption{The quantum circuit shown is for \(L=6\) and \(N=4\). The first four qubits from the top represent the system: two in \(|0\rangle\) and two prepared from \(|11\rangle\) into the antiferromagnetic cat \((|01\rangle-|10\rangle)/\sqrt{2}\), the $N=4$ case of Eq.~\ref{eq.zeta}. $M =2$ are the bottom two qubits initialised in the ground state found via exact diagonalisation and with $\eta_I = 1.107$. The circuit is divided into two parts. The first prepares the environment in the state obtained via diagonalisation and the system in the state of Eq.~\ref{eq.zeta}. The second implements two second-order Trotter steps, $e^{-i\hat H_X\tau/2}e^{-i\hat H_Z\tau}e^{-i\hat H_X\tau}e^{-i\hat H_Z\tau}e^{-i\hat H_X\tau/2}$, in which the adjacent half-steps of the transverse field between the two steps are merged into a single $R_x(2\phi)$ layer. The angles of rotation are given in Eq.~\ref{eq.angles}.}
    \label{fig: Toymodel}
\end{figure*}
    
\subsection{A Toy Quantum Circuit}
\label{sec.circuit}
As a visualisation tool for this MPS model, a toy model is worked through, including a circuit diagram, decompositions and angles. For a system of $L=6$ qubits, with $N=4$ and $M=2$, implementing the Hamiltonian defined in Eq.~\ref{eq.hamiltonian}. The time evolution is discretised into small steps $\tau$, where $\tau \ll t_{total}$, and the decomposition is applied iteratively. When an evaporation event occurs, it is captured by updating the gate set: the interaction gates corresponding to the system-environment boundary are shifted, and the departing system qubit is thereafter treated with environment interaction gates.

Since all $\hat\sigma^z\hat\sigma^z$ and $\hat\sigma^z$ terms mutually commute, $\hat H(t)=\hat H_X+\hat H_Z$ may be written, with
$\hat H_X=-\tfrac{g_{sys}}{2}\sum_{j\le N(t)}\hat\sigma^x_j-\tfrac{g_{env}}{2}\sum_{j>N(t)}\hat\sigma^x_j$
and $\hat H_Z$ containing all remaining terms, and use the second-order (Strang) splitting \cite{suzuki}
\begin{equation}
\hat U(\tau)=e^{-i\hat H_X\tau/2}\,e^{-i\hat H_Z\tau}\,e^{-i\hat H_X\tau/2}+\mathcal O(\tau^3).
\label{eq.strang}
\end{equation}
Each factor is a product of commuting local terms and so factorises exactly,
\begin{widetext}
\begin{align}
e^{-i\hat H_X\tau/2}&=\prod_{j=1}^{N(t)}e^{i g_{sys}\tau\hat\sigma^x_j/4}\prod_{j=N(t)+1}^{L}e^{i g_{env}\tau\hat\sigma^x_j/4},\label{eq.UX}\\
e^{-i\hat H_Z\tau}&=\prod_{j=1}^{N(t)-1}e^{iJ_{sys}\tau\hat\sigma^z_j\hat\sigma^z_{j+1}/4}\;
e^{ih\tau\hat\sigma^z_{N(t)}\hat\sigma^z_{N(t)+1}/4}
\prod_{j=N(t)+1}^{L-1}e^{iJ_{env}\tau\hat\sigma^z_j\hat\sigma^z_{j+1}/4}
\prod_{j=1}^{N(t)}e^{ih_z\tau\hat\sigma^z_j/2},\label{eq.UZ}
\end{align}
\end{widetext}
so the only Trotter error arises from $[\hat H_X,\hat H_Z]$, and the ordering of the gates within $e^{-i\hat H_Z\tau}$ is immaterial. This decomposition assumes the convention $\hbar = 1$.

Quantum computers are susceptible to various noise sources and gate errors (see, for example, \cite{knill}). Consequently, simplifying circuit implementation is imperative to optimise performance on current quantum hardware. Although the two-site gates commute, gates sharing a qubit cannot act simultaneously on hardware, so they are arranged in a brickwork: two-site operators first act upon all odd-labelled pairs, followed by a layer acting upon all even-labelled pairs. This structure maps directly onto standard one- and two-qubit rotation gates.

With $R_\alpha(\theta)=e^{-i\theta\hat\sigma^\alpha/2}$ and $R_{zz}(\theta)=e^{-i\theta\hat\sigma^z\hat\sigma^z/2}$, the gate angles are
\begin{align}
\phi_{sys,env}&=-\tfrac{\tau g_{sys,env}}{2},\nonumber\\
\theta_{sys,env}&=-\tfrac{\tau J_{sys,env}}{2},\nonumber\\
\theta_z&=-\tau h_z,\nonumber\\
\theta_h&=-\tfrac{\tau h}{2},
\label{eq.angles}
\end{align}
where $\phi$ is the half-step angle of Eq.~\ref{eq.UX}; adjacent half-steps of consecutive Trotter steps merge into a single $R_x(2\phi)$.

To initialise the environment: the ground state of $\hat{H}_{env}$ is directly diagonalised, yielding a weighted Bell-type state. This state can be prepared using a rotation-y gate in conjunction with standard entangling operations. The system comprises $4$ qubits, with the first two initialised in the $\ket{0}$ state and the remaining two in the antiferromagnetic cat $(\ket{01}-\ket{10})/\sqrt{2}$. This initialisation sequence is shown in the first section of Fig.~\ref{fig: Toymodel}. Subsequently, the figure illustrates a complete circuit for two Trotter steps; these steps are repeated for a total of $n$ iterations to simulate a sufficient duration of the system dynamics. For a full model compatible with quantum hardware, the entropy would be estimated from randomised measurements \cite{Elben2018,Brydges2019}; see also Ref.~\cite{CHOWDHURY2025117112} for a related circuit implementation.

\section{Criticality}
\label{app:criticality}

At $h_z=0$ the chain is critical at $J=2g$, i.e.\ at $\lambda\equiv J/2g=1$, where the single-particle gap $\Delta=|g-J/2|$ closes and the ground-state correlation length diverges. The main-text parameters place both chains in the disordered phase ($\lambda=\tfrac12$). Proximity to criticality could plausibly affect the entropy profile in two ways. First, it changes the quasiparticle dispersion and hence the rate at which entanglement spreads. Second, and more significantly, at criticality the environment's initial state $\ket{\phi_0}$ is no longer short-range correlated. A critical ground state carries entanglement growing logarithmically with subsystem size and is gap-less. Since $\hat H_{\mathrm{int}}$ and the longitudinal field act only on the system, the environment is a TFIM in both the integrable and chaotic protocols, so its proximity to criticality is well defined in both cases. The same is not true of the system. For $h_z\neq0$ the longitudinal field removes the critical point, so $\lambda_{\mathrm{sys}}=1$ is critical only in the integrable case.

\begin{figure*}
  \includegraphics[width=\textwidth]{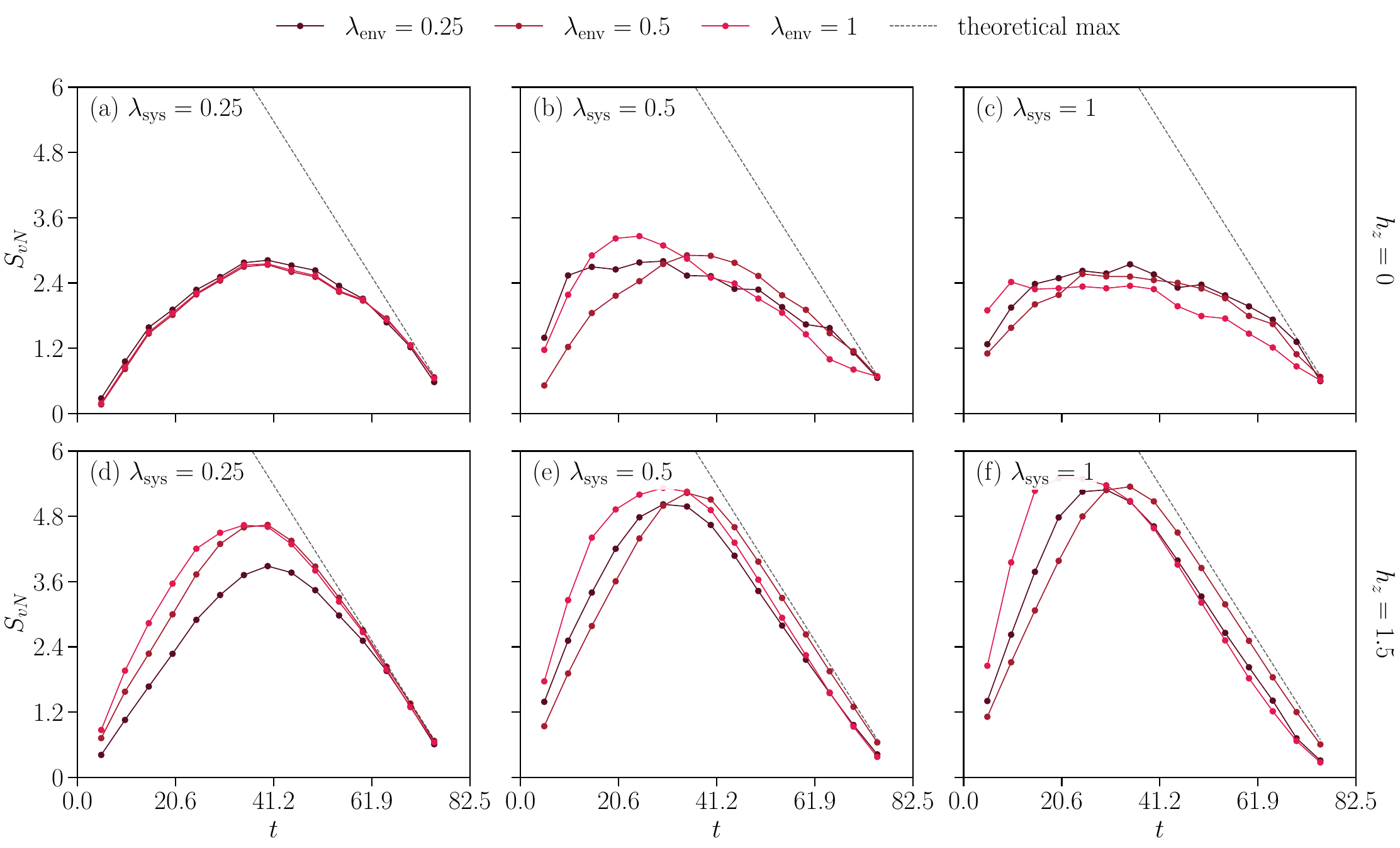}
  \caption{von Neumann entropy of the environment under the evaporation
  protocol as the system and environment approach the TFIM critical
  point $\lambda\equiv J/2g=1$. Columns:
  $\lambda_{\mathrm{sys}}=0.25$, $0.5$ (main-text value) and $1$
  [$(J,g)=(3,6),\,(3,3),\,(6,3)$]. Curves:
  $\lambda_{\mathrm{env}}=0.25$, $0.5$ and $1$
  [$(J_{\mathrm{env}},g_{\mathrm{env}})=(1.5,3),\,(1,1),\,(3,1.5)$].
  Top row: integrable ($h_z=0$); bottom row: chaotic ($h_z=1.5$), for
  which $\lambda_{\mathrm{sys}}=1$ is not critical. All other
  parameters are as in Fig.~\ref{fig:hzplot} ($N=15$, $M=100$, $T=5$, $h=3$). Markers are measurements made just before each
  evaporation event; the dashed line is the bound of Eq.~\ref{eq.bound S}.}
  \label{fig:criticality}
\end{figure*}

Changing $\lambda_{\mathrm{env}}$ also changes the environment's quasiparticle velocity. From Eq.~\ref{eq:vmax}, $v_{\mathrm{env}}=0.75$, $0.5$ and $1.5$ for $\lambda_{\mathrm{env}}=0.25$, $0.5$ and $1$, so the critical environment is also the fastest. The environment couplings visibly affect the early-time growth. In panels $(b)\text{ and }(c)$, curves reach a given entropy in the order $\lambda_{\mathrm{env}}=1$, $0.25$, $0.5$. This follows $v_{\mathrm{env}}$ rather than distance from the critical point: $\lambda_{\mathrm{env}}=0.25$ is further from criticality than $\lambda_{\mathrm{env}}=0.5$ yet grows faster. The exception is panel $(a)$, where in the chaotic case $\lambda_{\mathrm{env}}=0.25$ grows most slowly. In the integrable chain with a critical system (seen in $(c)$, on the top row) of Fig.~\ref{fig:criticality}, the critical environment also lies below the other curves on the descending branch.

None of these differences alter the conclusions of the main text. Across all nine combinations of $\lambda_{\mathrm{sys}}$ and $\lambda_{\mathrm{env}}$, the integrable chain peaks at $S_{vN}\approx 2.4$--$3.3$ and remains well below the bound of Eq.~\ref{eq.bound S} after the peak. The chaotic chain peaks at $S_{vN}\approx 3.9$--$5.5$, and its descending branch approaches the bound in every case. The largest variation within the chaotic chain occurs in $(a)$, where $\lambda_{\mathrm{sys}}=\lambda_{\mathrm{env}}=0.25$ peaks at $S_{vN}\approx 3.9$ against $\approx 4.6$ for the other environments, the lowest chaotic peak in this case still exceeds every integrable curve. The spread induced by the couplings, including tuning the system, the environment, or both to the critical point, is therefore smaller than the separation between the integrable and chaotic chains. The environment couplings modify the rate of entropy growth, predominantly through the quasiparticle velocity, but proximity to criticality does not account for the difference in how closely the two chains approach the Page bound.

 \newpage
 
\bibliographystyle{apsrev4-2}
\bibliography{references}
\end{document}